\documentclass[10pt]{iopart}
\usepackage[pdftex]{hyperref}
\usepackage{mathptmx}
\usepackage[squaren]{SIunits}
\usepackage{graphicx}
\usepackage{caption}
\usepackage{cleveref}
\usepackage{xcolor}
\usepackage{cite}
\usepackage{amssymb}

\crefname{figure}{figure}{figures}
\Crefname{figure}{Figure}{Figures}

\graphicspath{ 
{./images/}
}

\begin{document}

\title[Micro Fresnel Mirror Array]{Micro Fresnel Mirror Array with Individual Mirror Control}

\author{Binal Poyyathuruthy Bruno$^{1}$, Robert Sch\"utze$^{1}$, Ruediger Grunwald$^{2}$ and Ulrike Wallrabe$^{1,}$}

\address{%
$^{1}$ \quad Laboratory for Microactuators, IMTEK - Department of Microsystems Engineering, University of Freiburg, Georges-K\"ohler-Allee 102, 79110, Freiburg, Germany\\
$^{2}$ \quad Max-Born-Institute for Nonlinear Optics and Short Pulse Spectroscopy, Berlin, Germany}
\ead{wallrabe@imtek.uni-freiburg.de}
\vspace{10pt}
\begin{indented}
\item[]October 2019
\end{indented}

\begin{abstract}
We present the design and fabrication of a miniaturized array of piezoelectrically actuated high speed Fresnel mirrors with individual mirror control. These Fresnel mirrors can be used to generate propagation invariant and self-healing interference patterns. The mirrors are actuated using piezobimorph actuators, and the consequent change of the tilting angle of the mirrors changes the fringe spacing of the interference pattern generated. The array consists of four Fresnel mirrors each having an area of \mbox{2x2 $mm^2$} arranged in a 2x2 configuration. The device, optimized using FEM simulations, is able to achieve maximum mirror deflections of \mbox{15 mrad}, and has a resonance frequency of \mbox{28 kHz}. 
\end{abstract}

%
%
\submitto{\SMS}
%
\maketitle
%
\ioptwocol

\section{Introduction}

A Fresnel mirror consists of two mirror planes inclined at a small angle to each other. It generates an interference pattern in the reflected light \cite{Fresnel1868} and proves the wave nature of light \cite{waveProof}. When illuminated with a coherent light source, the two mirrors split the light into two coherent overlapping light sheets with planar wavefronts, thereby generating an interference pattern along the optical axis. The spatial frequency of the resulting pattern is propagation invariant and self-healing, similar to the central maxima of a Bessel beam. If the pattern is obstructed by an object, the interference fringes will form again further down the optical axis. The size of the interference pattern depends on the tilt of the mirrors and the wavelength of the illuminating light. 

Fresnel mirrors can, for example, be used to fabricate linear optical gratings directly into the photoresist without the use of a photomask \cite{Malag1980}. It was shown that an adaptive Fresnel mirror can be used in adaptive pulse autocorrelation \cite{Treffer2016}, and for direct nano-machining of sub-wavelength structures on silicon or dielectric substrates by laser-induced periodic surface structures (LIPSS) \cite{Brunne2012,Treffer2013}. LIPSS are formed by the interference of incident and scattered light pulses on the surface of the material \cite{Das2010}. Fresnel mirrors deliver a sharp interference pattern with minimal optical dispersion, thereby creating coherently linked nano-structures. The interference pattern generated using a Fresnel mirror can also be used in 3D scanners, where the pattern is projected onto an object and the distortion of lines is captured by a camera to measure the surface topology \cite{3Dtopo}. 
 
The first MEMS Fresnel mirror was fabricated by Oka et. al \cite{Oka2011} for a microfluidic diffusion sensor. Brunne et. al.\cite{Brunne2013} developed the first Fresnel mirror with adjustable tilting angle with an aperture of \mbox{5 mm}. The design consisted of two mirror segments, which were attached to a silicone (PDMS) base layer using silicone glue, which was in turn glued to a piezoceramic disc. The fabrication process in \cite{Brunne2013} relied heavily on rapid laser prototyping, which introduced non-uniformities in the process, and the use of silicone glue reduced the repeatability of the process.

In this paper, we present the miniaturization and optimization of a 2x2 array of adaptive Fresnel mirrors. We also demonstrate the individual tunability and study the cross-talk between the mirrors. Such an array may find application in studying discretely spatially resolved autocorrelation and in multichannel line foci for parallel material processing. In contrast to previously reported studies on 2x2 circular axicons arrays\cite{Treffer2017}, the optical symmetry of Fresnel mirror arrays is reduced by one dimension. Resulting interference characteristics are essentially different. In \cite{FresnelISOT}, we have already reported on the initial fabrication and measurement results of the array. Here, the simulation and fabrication is further refined, and more detailed characterizations are done. The paper is organized as follows: in \cref{sec:Design}, we describe the working principle and the design optimization of the device. The fabrication process discussed in \cref{sec:Fab} is followed by the mechanical and optical characterization, and optimization of the mirror deflection in \cref{sec:Meas}. The individual control of the array elements and the cross-talk effects are discussed in \cref{sec:Ind}. The results are summarized, and further improvements are discussed in \cref{sec:Summ}.

\section{Design and Simulation}
\label{sec:Design}
\begin{figure*}[ht!]
	\centering
	\def\svgwidth{1\textwidth}
	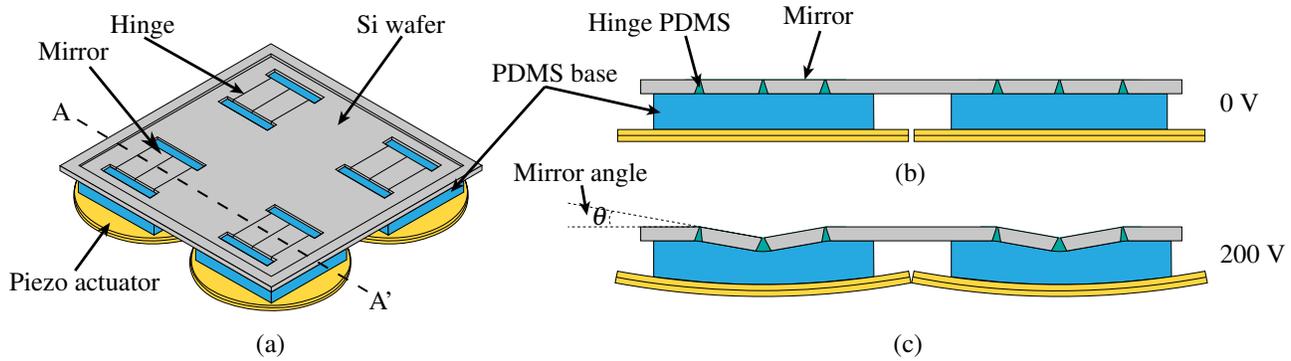
	\caption{(a) 3D model of 2x2 array of \mbox{2x2 mm\textsuperscript{2}} Fresnel mirrors showing different components of the mirror. Working principle of the tunable Fresnel mirror, (b) showing cross-section (A-A') of the mirrors at \mbox{0 V} and (c) at \mbox{-200 V}.}
	\label{fig:DeviceCross}
\end{figure*}

The array of Fresnel mirrors presented here consists of four Fresnel mirrors, each with an aperture of \mbox{2 mm}, arranged in a 2x2 configuration as can be seen in \cref{fig:DeviceCross} (a). A single Fresnel mirror consists of two mirror segments that meet at the center line of the device. The mirrors are attached to a silicone (PDMS) base layer, which is in turn attached to a piezoceramic disc (\cref{fig:DeviceCross} (b)). Upon actuation, the piezo discs undergo a spherical deflection, which act as a pressure load on the PDMS layer and thereby pulling the mirrors downwards (\cref{fig:DeviceCross} (c)). The mirrors are made from silicon wafer coated with aluminium, making it stiff and thereby able to maintain their planar surface. 

\subsection{Optimization by FEM-simulation}

The tunable range of the mirror angle depends on the geometry of the hinge, the thickness of the silicon wafer and the PDMS base layer, the Young’s modulus of the PDMS at the hinges and the base layer, and the piezoelectric material used. A \mbox{2 mm} Fresnel mirror with quarter symmetry was simulated in \textsl{COMSOL Multiphysics} to design for a high tuning range. To facilitate the tilting movement of the mirrors, the hinge geometry was designed to be trapezoidal with a hinge width as narrow as possible on the mirror side. The dimensions of the hinge are limited by the new fabrication process based on KOH etching. The individual mirrors are made from a silicon wafer and are actuated using circular piezo bimorphs with a thickness of \mbox{240 $\mu$m}. A separate material study was conducted to determine the best suited piezomaterial for the device \cite{PZTCompare}. The diameter of the individual actuators was chosen using a parametric sweep with a tradeoff between the achievable mirror angle and the pitch of the array. The Young's modulus and the Poisson ratio of the silicone were taken from the materials library of \textsl{COMSOL}. A simple linear model was sufficient for the simulation since the expected strain levels were small, in the range of $<10\%$ \cite{Schneider2008}. The strain-charge form of piezoelectric simulation was used, which operates on solid mechanics and electrostatics physics in \textit{COMSOL}. The model is considered fixed only on the outer boundary of the mirror frame and everything else is considered free to move.

\begin{figure*}[ht!]
	\centering
	\def\svgwidth{1\textwidth}
	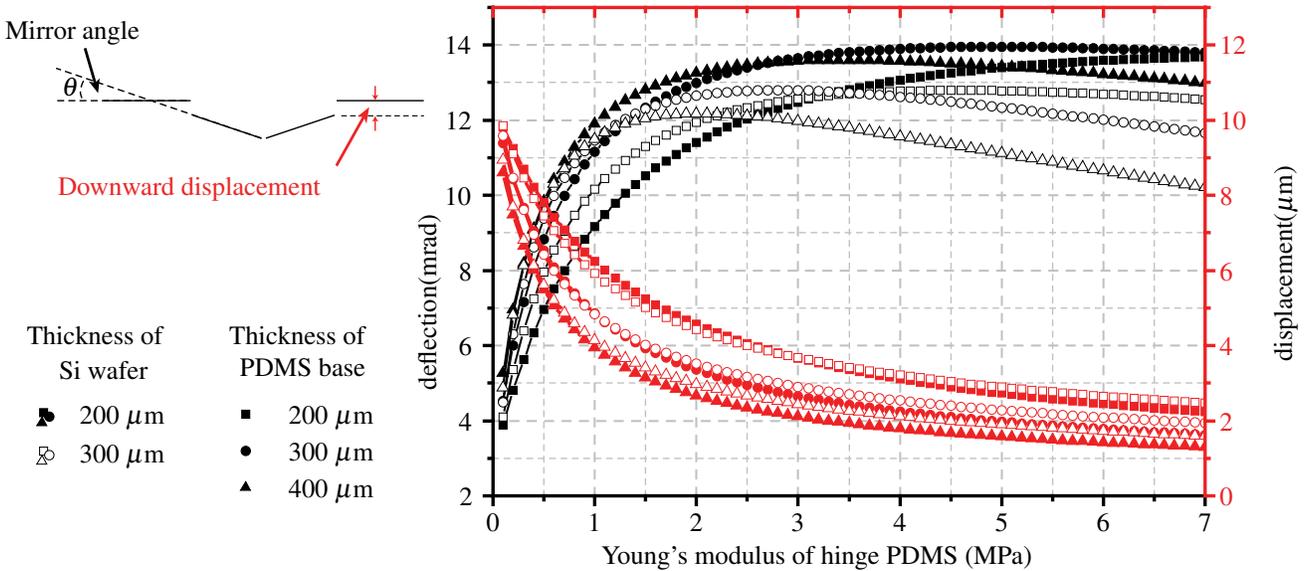
	\caption{Simulated mirror angle and downward displacement of the Fresnel mirror as a function of the Young's modulus of the PDMS in the hinges and at varying thickness of silicon wafer and the PDMS base.}
	\label{fig:HingeSim}
\end{figure*}

Furthermore, when the mirrors are pulled downwards, two situations arise according to the stiffness of the hinge. A high stiffness suppresses the rotation around the hinge resulting in a low tilting angle, whereas a low stiffness introduces an out of plane translational displacement, which dominates the rotation of the hinge. The variation in the tilting angle and the downward deflection with regard to the Young's modulus of the hinge PDMS for varying thickness of the PDMS base layer and the Si wafer is shown in \cref{fig:HingeSim}. The maximum mirror angle is achieved when the hinge has a stiffness of \mbox{5 MPa} with a \mbox{300 $\mu$m} thick base layer and a \mbox{200 $\mu$m} thick Si wafer. Nevertheless, a PDMS with a Young's modulus of \mbox{1.5 MPa} was chosen for fabrication because of the availability of the material. With this configuration, the maximum achievable mirror angle was simulated to be in the range of 12 to \mbox{13 mrad}, which is 92\% of the value at the maximum value at \mbox{5 MPa}, and the downward displacement of the mirrors amounts to \mbox{4 $\mu$m}. All simulations were done at \mbox{200 V} with a voltage correction factor to adapt the material parameter for the piezoelectric material used \cite{PZTVoltScale}.

\section{Fabrication}
\label{sec:Fab}
\begin{figure*}[ht!]
	\centering
	\def\svgwidth{1\textwidth}
	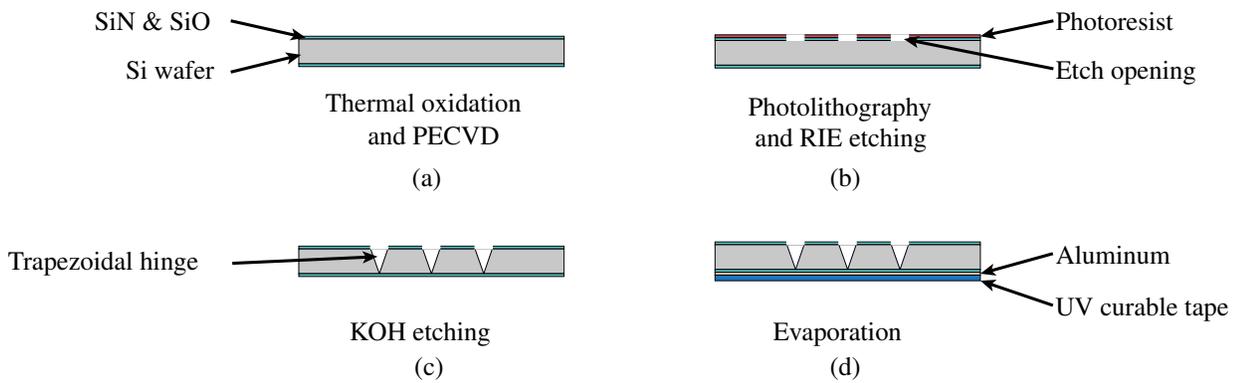
	\caption{Cleanroom process steps for manufacturing the mirror hinges. (a) SiO and SiN deposited by thermal oxidation and PECVD process respectively. (b) Etch opening for the hinges made by photolithography and RIE etching of the oxide and nitride layers. (c) Trapezoidal hinge made by the directional KOH etching of Si wafer. (d) Evaporated aluminum acting as the mirror layer and the UV curable tape for wafer protection.}
	\label{fig:Fab0}
\end{figure*}

The fabrication process starts with the manufacturing of the mirror hinges on a \mbox{200 $\mu$m} thick Silicon wafer. A \mbox{400 nm} layer of SiO and \mbox{110 nm} of SiN layer is deposited on the silicon wafer using thermal oxidation and LPCVD process respectively (\cref{fig:Fab0} (a)). The oxide and nitride layers act as an etch mask for the KOH etching process and are opened up on the back side using reactive ion etching (\cref{fig:Fab0} (b)). The wafer is etched using KOH to obtain a trapezoidal hinge geometry (\cref{fig:Fab0} (c)). A \mbox{300 nm} thick aluminum layer is deposited on the front side of the wafer to act as the mirror layer. The wafer is covered with a layer of UV curable tape to protect the mirror surface during subsequent processing (\cref{fig:Fab0} (d)), and individual 4x4 mirror array chips are seperated using UV laser. 
 
\begin{figure*}[ht!]
	\centering
	\def\svgwidth{1\textwidth}
	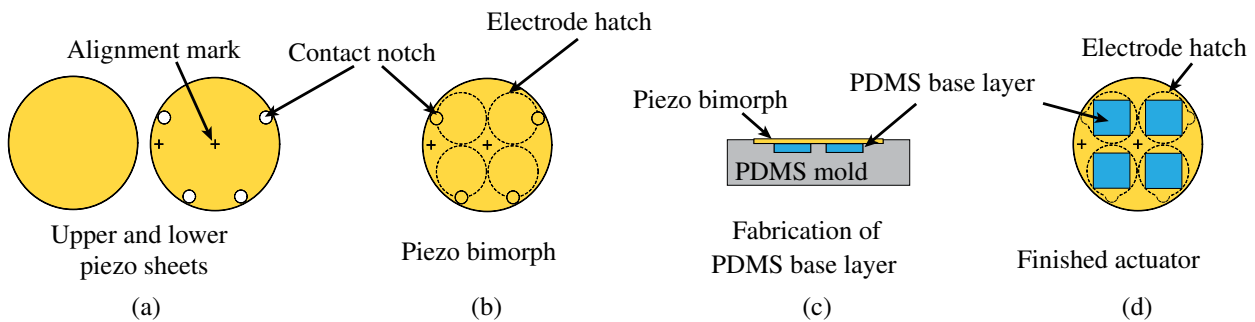
	\caption{Fabrication of the actuator from two seperate piezo sheets (a) glued together to form a bimorph (b). (c) Fabrication of the PDMS base layer using a negative mold. (d) Finished actuator with base layer ready for assembly.}
	\label{fig:Fab1}
\end{figure*}

The piezo actuator is fabricated from two \mbox{120 $\mu$m} thick sheets of PZT from \textsl{Ekulit GmbH}. The upper and lower piezo sheets are structured into a circular shape using a UV laser (\textsl{Trumark 6330} from \textsl{Trumpf Laser GmbH}) (\cref{fig:Fab1} (a)). The two piezo sheets are cleaned in an ultrasonic bath and glued together using high-temperature epoxy (\textsl{HTG-240} from \textsl{Resoltech}) to form the piezo bimorph. Afterward, the bimorph is aligned in the UV laser, and the electrodes for each mirror in the array are separated by hatching, as shown in \cref{fig:Fab1} (b). The piezo actuator is coated with an adhesion promoter (Dow Corning 92-023) and the PDMS base layer (\textsl{RTV23} from \textsl{Neukasil}) with Young’s modulus of \mbox{200 kPa} \cite{RTV23}, is fabricated directly on the bimorphs using a negative mold (\cref{fig:Fab1} (c) \& (d)). 
 
\begin{figure*}[ht!]
	\centering
	\def\svgwidth{1\textwidth}
	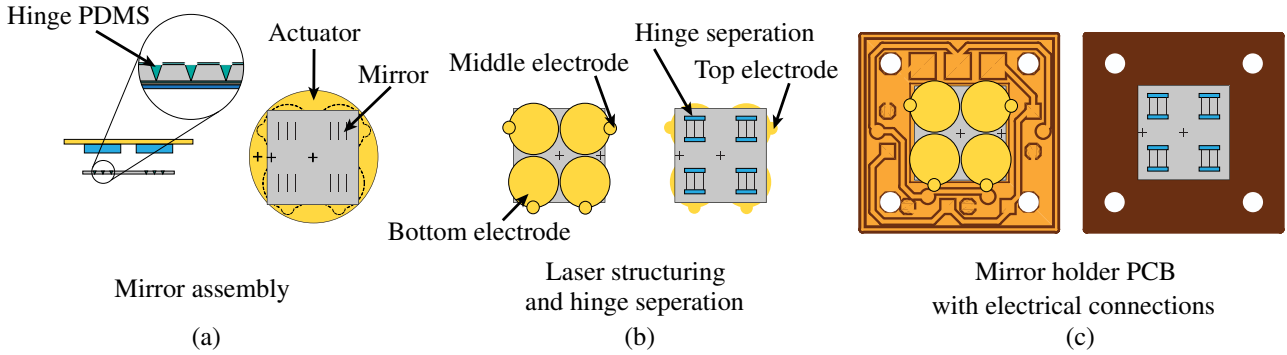
	\caption{(a) Filling the hinges with PDMS and assembling the actuator. (b) Decoupling of the individual actuators and seperation of mirror hinges. (c) Finished mirror array glued on the mirror PCB for electrical connections.}
	\label{fig:Fab2}
\end{figure*}

The individual mirror array chip is diced from the Si wafer again using the UV laser and is cleaned in an ultrasonic bath using deionized water. The backside of the hinges is coated with an adhesion promoter (\textsl{Dow Corning 92-023}), and the hinges are filled with PDMS (\textsl{RTV615} from \textsl{Momentive Performance Materials}). The nitride and oxide layers underneath the aluminum layer prevent the creeping of the PDMS to the front side of the mirror. The mirror chip is placed on a vacuum chuck, and the actuator is aligned and attached to the backside of the mirror while the PDMS is being cured (\cref{fig:Fab2} (a)). After curing, the individual actuators are decoupled and separated by laser structuring from the backside, and the mirrors are separated from the front side, making them free to move around the hinges (\cref{fig:Fab2} (b)). The UV tape on the front side is exposed to UV light and peeled off to reveal the mirror surface. The finished mirror array is carefully aligned and glued to a laser structured FR4 PCB-substrate (\cref{fig:Fab2} (c)), and the electrodes are soldered to the PCB using a \mbox{100 $\mu$m} thick copper wire. 

\section{Optimization of the deflection angle}
\label{sec:Meas}
To analyze the deflection characteristics, the device was mounted on a custom mount, which also contained an electrical protection circuit that limited the voltage against the direction of polarization to within 33\% to 50\% of the coercive field strength and thereby prevented the depolarization of the piezo sheets \cite{PZTCompare}. The mirrors were actuated using a \mbox{1 Hz} sinusoidal signal, and the deflection of the mirrors was measured using a profilometer equipped with a confocal distance sensor. The angle between the device plane and the mirror was evaluated from the quasi-statically measured mirror profile.

\begin{figure*}[ht!]
	\centering
	\def\svgwidth{1\textwidth}
	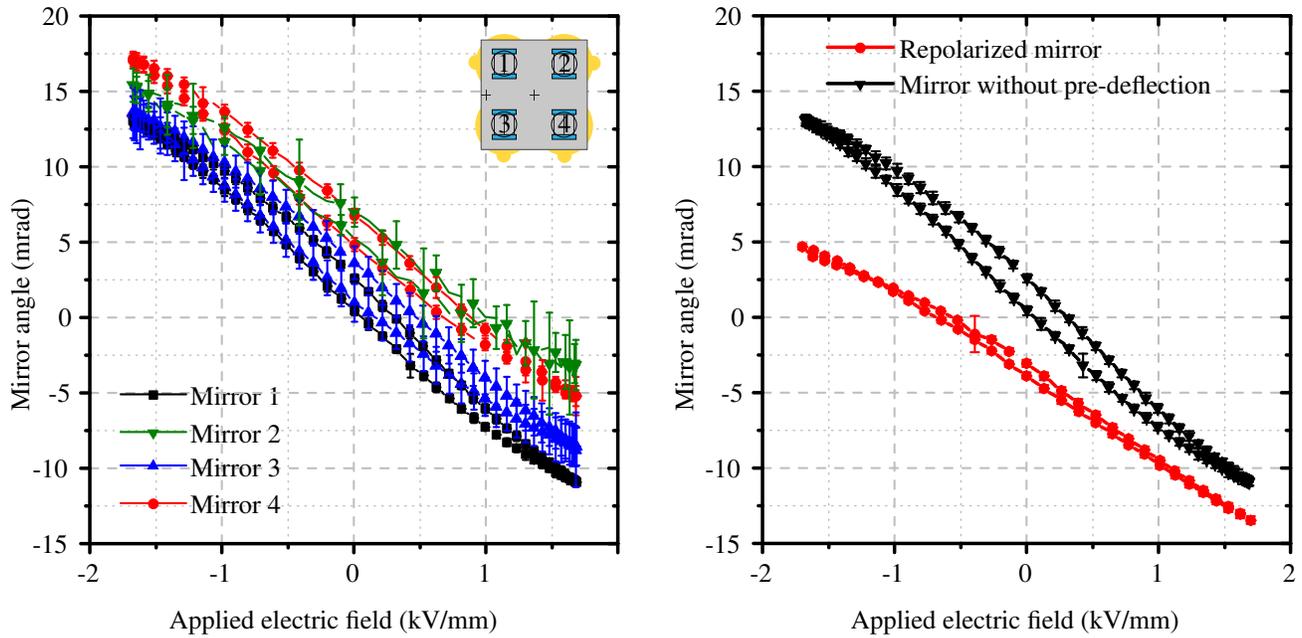
	\caption{(a) Tilt angle of the mirrors in the array as a function of the applied electric field. (b) Induced negative predeflection by depolarizing the upper piezo sheet and repolarizing it after the fabrication of the mirrors. The depolarized mirror also shows reduced hysteresis.}
	\label{fig:mirrorAngle}
\end{figure*}

\Cref{fig:mirrorAngle} (a) shows the tilt angle as a function of the applied electric field, with a negative tilt angle denoting the mirror deflecting downwards. The maximum deflection range reached \mbox{22 mrad}, well near the simulated results and the maximum downward displacement of the mirrors was \mbox{2 $\mu$m}. However, each mirror showed a different, undesired pre-deflection. Typically, only negative mirror angles are required for the application, which reduces the deflection range of the mirrors in the symmetric case by 50\%. To overcome this problem, a pre-deflection can be defined by depolarizing the upper piezo sheet before gluing and finally re-polarizing the piezo material by applying an electric field higher than the coercive field strength of the material~\cite{Act2018PI,Act2018Florian}. This creates a remanent strain in the piezo bimorphs which induces a pre-deflection of the mirrors in the negative direction. This was proved as shown in \cref{fig:mirrorAngle} (b) and the depolarized mirror also shows reduced hysteresis. The maximum deflection range can also be increased further by changing the operating electric field range. It was previously observed that the electric field applied against the direction of polarization can be increased up to 95\% of the coercive field while operating using symmetric cycles, and up to 50\% while driven only with negative cycles \cite{PZTCompare}. The electrical protection circuit was modified to limit the electric field against the direction of polarization to 50\%.

\begin{figure*}[ht!]
	\centering
	\def\svgwidth{1\textwidth}
	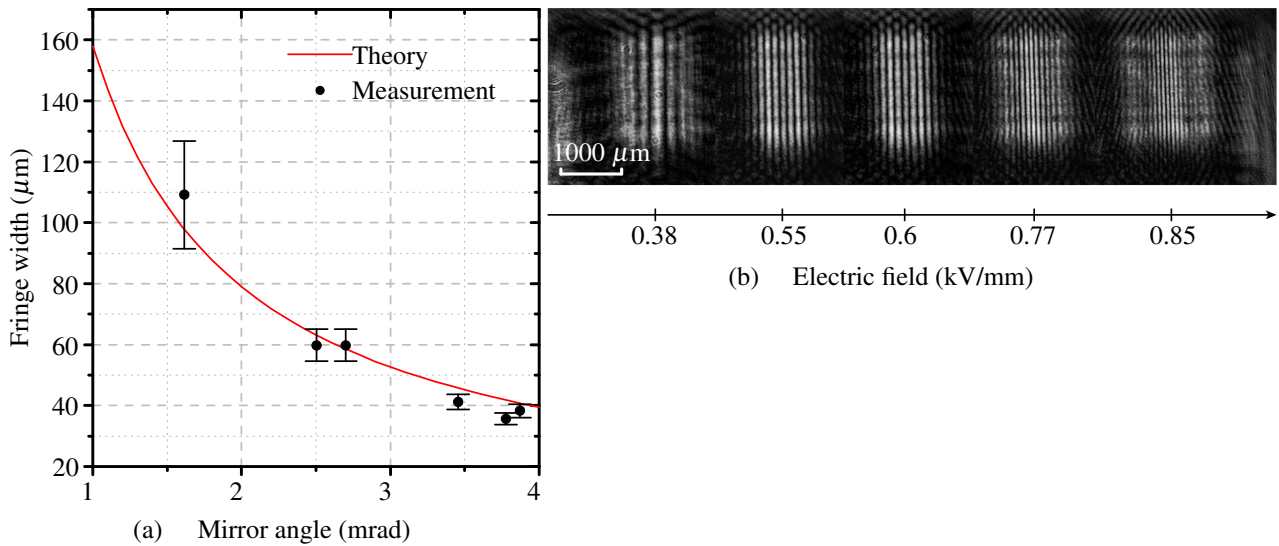
	\caption{(a) The fringe spacing measured for different mirror angles compared to the theoretical values. (b) The fringe pattern at different mirror angle captured using a CMOS camera.}
	\label{fig:opticalMeas}
\end{figure*}

To also characterize the tunable deflection of the mirror with its optical function, it was illuminated using a low power laser \mbox{($\lambda = 632 nm$)} and the interference pattern generated was imaged using a CMOS camera at a distance of \mbox{150 mm} from the mirror. A DC signal was applied to the mirror and after waiting several minutes to account for the creeping, the interference pattern was recorded. \Cref{fig:opticalMeas} (b) shows the interference pattern at different electric fields. The geometric line spacing between the fringes is given by \mbox{$\Delta x = {\lambda}/{4\beta}$}, where $\lambda$ is the wavelength of the incident beam and $\beta$ is the mirror angle \cite{Brunne2013}. \Cref{fig:opticalMeas} (a) compares the theoretical fringe spacing as expected from the simulations to the one evaluated from the captured images. The measured fringe spacing corresponds well to the prediction with a deviation of less than 4\%.

\section{Individual mirror control and cross-talk}
\label{sec:Ind}
\begin{figure*}[ht!]
	\centering
	\def\svgwidth{1\textwidth}
	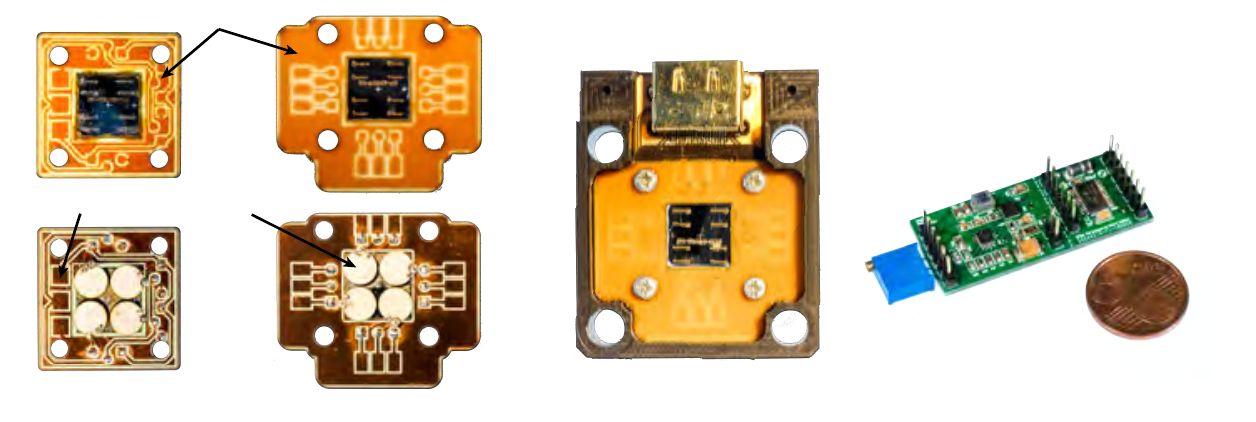
	\caption{Finished mirror arrays parallel control (a) and individual control (b) configurations. Mirror mount (c) and the control circuit (d) for individually controllable mirror array.}
	\label{fig:MirrorMount}
\end{figure*}

Even though we were able to increase the range of the usable angle from \mbox{10 mrad} to \mbox{15 mrad}, the dissimilarity amoung the mirrors in the same array still remains. This might be due to the small mass of the mirror, together with the flexibile hinge and PDMS base, making the mirror angle sensitive to various ambiguities arising during the rapid prototyping process. To compensate for this, the mirror PCB was modified to control each mirror individually (\cref{fig:MirrorMount} (b)). In order to do so, a control circuit was built using a boost converter IC (\textsl{HV9150} from \textsl{Microchip Technology Inc.}) and quad channel opamp (\textsl{HV264} from \textsl{Microchip Technology Inc.}) (\cref{fig:MirrorMount} (d)). The circuit is capable to provide four different control signals with voltages up to \mbox{200 V} and with maximum operating frequency of \mbox{20 kHz}.

\begin{figure*}[ht!]
	\centering
	\def\svgwidth{1\textwidth}
	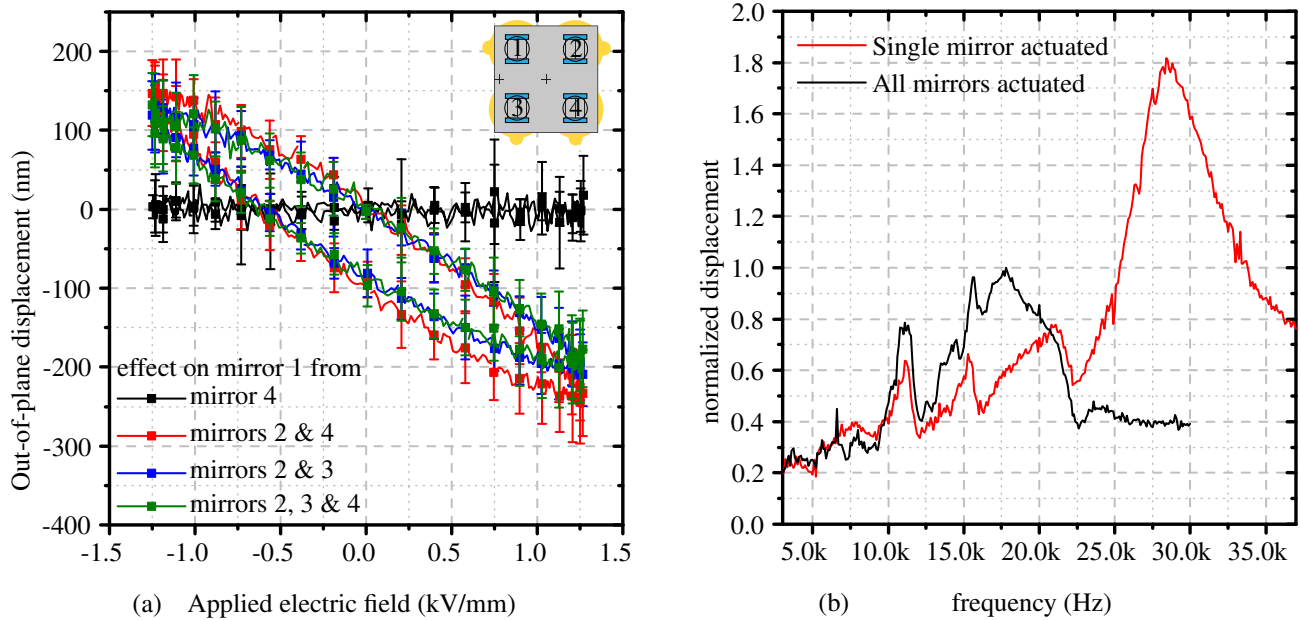
	\caption{(a) Out-of-plane displacement measured on a stationary mirror due to the mechanical cross-talk from the rest of the mirrors in the array. (b) Frequency response of a single Fresnel mirror in the array and frequency response when all mirrors are actuated simultaneously, normalized to the maximum displacement at simultaneous actuation.}
	\label{fig:crossRes}
\end{figure*}

The mechanical cross-talk between the mirrors at quasi-static operation was measured separately by actuating the mirrors in different configurations and measuring the resulting displacement in an unactuated mirror using the profilometer. The actuation of adjacent mirrors will create an out-of-plane displacement in the unactuated mirror, without changing the mirror angle. \Cref{fig:crossRes} (a) shows the passive out-of-plane displacement of mirror one in the array as a result of different actuating conditions of the other mirrors. The large error bars are mainly the result of the error from the confocal sensor \mbox{($\pm$50 nm)} used for measuring the deflection. The cross-talk was observed mainly from the horizontally adjacent mirror, with the diagonal and vertical neighboring mirrors showing no significant effect. Even though the horizontal and vertical mirrors are in similar positions relative to one another, the cross-talk is asymmetric because the horizontal boundary is a hinge and the vertical boundary is a separation from the mirror frame.

For evaluating the dynamic performance of the device, the mirrors were actuated using a sinusoidal signal with an amplitude of \mbox{10 V} at different frequencies and the deflection of the mirrors were measured using a triangulation sensor. \Cref{fig:crossRes} (b) shows the frequency response of the mirror at two different actuations. A resonance peak at \mbox{28 kHz} was observed when only one mirror was actuated. However, when all the mirrors were actuated simultaneously, the frequency of the peak mirror angle was reduced to \mbox{18 kHz}. All other actuation combinations (2 and 3 mirrors actuated simultaneously) had resonance peaks in between. The change in the resonance possibly arises due to the mechanical cross-talk between the mirrors and electrical cross-talk between the piezo actuators and needs further investigation.

\section{Summary and conclusion}
\label{sec:Summ}
In this paper, we presented the design and fabrication of a 2x2 array of Fresnel mirrors with an aperture of 2 mm each. The device was optimized using finite element simulations predicting a maximum mirror deflection of \mbox{12 mrad}. The fabrication process was developed and relies on KOH etching, PDMS processing, and laser cutting. 

The mechanical characterization of the fabricated device yielded a similar range of tilt angle as the simulation. As only negative angles are required for applications, we were able to increase the maximum tilt angle in the negative direction by up to 50\% by introducing a defined pre-deflection by means of depolarizing and re-polarizing the upper piezosheet. 

However, the mirrors in the array showed different pre-deflections, which is undesired as the mirrors need to be actuated synchronously. This proved to be challenging to eliminate entirely. Hence, each mirror needed to be addressed according to its own voltage dependence. For this purpose, we built an individually controllable array with accompanying electronics. By controlling each mirror individually, we were also able to measure the mechanical cross-talk between the mirrors in the array. The observed out-of-plane displacement of an unactuated mirror due to the cross-talk was in the range of \mbox{150 nm}, which is less than $\lambda/4$. 

We found that a single Fresnel mirror showed fast switching speeds with resonance frequency of \mbox{28 kHz}. However, the resonance frequency was reduced when operated along with the other mirrors in the array. 

The optical characterization of the mirrors was performed by geometrically estimating the line width of the interference pattern at different mirror angle. We verified the estimation by measuring the intensity distribution in the overlapping zone using a CMOS camera and found it to be in good agreement with the theoretical predictions.

To conclude, we were able to miniaturize the tunable Fresnel mirror and fabricate it in a 2x2 array. The optical and mechanical characterization showed good agreements with the simulations. The different pre-deflections in the mirrors could be further reduced by optimizing the tolerances in the fabrication process. We observed that a single Fresnel mirror had a resonance frequency of \mbox{28 kHz}. However, further investigation is needed to determine the coupling effects that cause the change in the resonance frequency at different operating modes.

\section*{Acknowledgments}
This work was supported by Deutsche Forschungsgemeinschaft (DFG) within grant no. WA1657/3-2. The costs to publish in open access was covered by Deutsche Forschungsgemeinschaft (DFG).

\section*{References}
\bibliographystyle{unsrt}
\bibliography{fresnelMirrorBib}

\end{document}